\documentclass[a4paper]{PoS}

\usepackage{amsmath}
\usepackage{cite}
\usepackage{subcaption}

\newcommand{\be}{\begin{equation}}
\newcommand{\ee}{\end{equation}}
\newcommand{\bea}{\begin{eqnarray}}
\newcommand{\eea}{\end{eqnarray}}

\usepackage{array}
\newcolumntype{C}{>{$}c<{$}}
\newcolumntype{L}{>{$}l<{$}}
\title{Lepton anomalous magnetic moments in Lattice QCD+QED}

\ShortTitle{Lepton anomalous magnetic moments}

\author{\speaker{D.~Giusti}$^{(a,b)}$ and S.~Simula$^{(b)}$

\\

\it $^{(a)}$ Dipartimento di Matematica e Fisica, Universit\`a  degli Studi Roma Tre, Rome, Italy.

\it $^{(b)}$ Istituto Nazionale di Fisica Nucleare, Sezione di Roma Tre, Rome, Italy.

\vspace{1.0cm}

Email: \email{davide.giusti@uniroma3.it}, \email{silvano.simula@roma3.infn.it}

}

\abstract{We present a lattice calculation of the Hadronic Vacuum Polarization (HVP) contribution to the anomalous magnetic moments of the electron, $a_e^{\rm HVP}$, the muon, $a_\mu^{\rm HVP}$, and the tau, $a_\tau^{\rm HVP}$, including both the isospin-symmetric QCD term and the leading-order strong and electromagnetic isospin-breaking corrections. Moreover, the contribution to $a_\mu^{\rm HVP}$ not covered by the MUonE experimen, $a_{MUonE}^{\rm HVP}$, is provided. We get $a_e^{\rm HVP} = 185.8~(4.2) \cdot 10^{-14}$, $a_\mu^{\rm HVP} = 692.1~(16.3) \cdot 10^{-10}$, $a_\tau^{\rm HVP} = 335.9~(6.9) \cdot 10^{-8}$ and $a_{MUonE}^{\rm HVP} = 91.6~(2.0) \cdot 10^{-10}$. Our results are obtained in the quenched-QED approximation using the QCD gauge configurations generated by the European (now Extended) Twisted Mass Collaboration (ETMC) with $N_f=2+1+1$ dynamical quarks, at three values of the lattice spacing varying from $0.089$ to $0.062$ fm, at several values of the lattice spatial size ($L \simeq 1.8 \div 3.5$ fm)  and with pion masses in the range between $\simeq 220$ and $\simeq 490$ MeV.}

\FullConference{37th International Symposium on Lattice Field Theory - Lattice2019\\
		16-22 June 2019\\
		Wuhan, China}

\begin{document}

\section{Introduction}
\label{sec:intro}

QCD simulations on the lattice represent nowadays the most promising tool for an accurate determination of the hadronic contributions to various electroweak observables directly from first principles. 
Here we consider the case of the anomalous magnetic moments of the three charged leptons of the Standard Model (SM): electron, $a_e$, muon, $a_\mu$, and tau, $a_\tau$.

Both $a_e$ and $a_\mu$ have been determined precisely from experiments:
\bea
    \label{eq:ae_exp}
        a_e^{exp} & = & 1~159~652~180.73 ~ (0.28) \cdot 10^{-12} ~~ [0.24~\mbox{ppb}] \qquad \mbox{\cite{Hanneke:2008tm}} ~ , ~ \\
    \label{eq:amu_exp}
    a_\mu^{exp} & = & 1~165~920~9.1 ~ (6.3) \cdot 10^{-10} ~ \qquad [0.54~\mbox{ppm}] \qquad \mbox{\cite{Bennett:2006fi}} ~ ,   
\eea
while the short lifetime of the tau ($\simeq 2.9 \cdot 10^{-13} \, \mbox{s}$) makes quite difficult a precise experimental determination of $a_\tau$. 

Within the SM the lepton anomalous magnetic moment is given by the sum of three terms, representing  the QED, the hadronic and the weak contributions. 
Taking the updated (five loops) QED contribution from Ref.~\cite{Aoyama:2017uqe} and the estimates of the hadronic and weak terms from Ref.~\cite{Jegerlehner:2017zsb} in the case of the electron and from Ref.~\cite{Davier:2019can} in the case of the muon, one finds the following deviations (anomalies) from the SM expectations:
\bea
    \label{eq:delta_ae_exp_SM}
            a_e^{exp} - a_e^{SM} & = & -1.30 ~ (0.77) \cdot 10^{-12} ~~ [1.7 \sigma] \qquad \mbox{\cite{Aoyama:2017uqe}} ~ , ~ \\
    \label{eq:delta_amu_exp_SM}
    a_\mu^{exp} - a_\mu^{SM} & = & +26.1 ~ (7.9) \cdot 10^{-10} \quad [3.3 \sigma] \qquad \mbox{\cite{Davier:2019can}} ~ .   
\eea
Recently a new determination of the fine structure constant $\alpha_{em}$ from atomic caesium~\cite{Parker:2018vye} leads to a more precise value of the electron anomaly
\be
    \label{eq:delta_amu_exp_SM_new}
    a_e^{exp} - a_e^{SM} = -0.88 ~ (0.36) \cdot 10^{-12} ~~ [2.4 \sigma] \qquad \mbox{\cite{Parker:2018vye}} ~ .
\ee
Notice the opposite signs of the electron and muon anomalies~\cite{Davoudiasl:2018fbb}.

The main sources of the uncertainty for the anomalies are the experiment and the QED calculation for $a_e$, while they are the experiment and the hadronic contribution for $a_\mu$.
New experiments at Fermilab (E989)~\cite{Logashenko:2015xab} and J-PARC (E34)~\cite{Otani:2015lra} aim at a fourfold reduction of the experimental uncertainty for the muon and important improvements in the precision are expected to come from the Harvard group for the electron.
On the theoretical side a significative reduction of the uncertainty is required for the QED contribution to $a_e$ and for the HVP and light-by-light contributions to $a_\mu$.

Nowadays the most accurate predictions for the HVP contribution to the muon, $a_\mu^{\rm HVP}$, come from the use of dispersion relations, which relate the HVP function to the experimental cross section data for $e^+ e^-$ annihilation into hadrons (see Ref.~\cite{Davier:2019can} and therein quoted).
In recent years lattice QCD calculations of $a_\mu^{\rm HVP}$ (see Refs.~\cite{Miura:2019xtd,Guelpers} for recent reviews) have made an impressive progress and they can provide a completely independent cross-check from first principles.

In this contribution we present our lattice determinations of the HVP contribution to the lepton anomalous magnetic moment $a_\ell^{\rm HVP}$ for $\ell = e, \mu, \tau$, using the QCD gauge configurations generated by ETMC with $N_f=2+1+1$ dynamical quarks, at three values of the lattice spacing varying from $0.089$ to $0.062$ fm, at several values of the lattice spatial size ($L \simeq 1.8 \div 3.5$ fm) and with pion masses in the range between $\simeq 220$ and $\simeq 490$ MeV.
Details concerning the $17$ ETMC gauge ensembles can be found in Table 1 of Ref.~\cite{Giusti:2018mdh}.

The hadronic quantities $a_\ell^{\rm HVP}$ are calculated including both the leading order (LO) generated by QCD effects in the isospin symmetric limit, contributing to order ${\cal{O}}(\alpha_{em}^2)$, and the next-to-leading order (NLO) corresponding to electromagnetic (em) and strong isospin-breaking (IB) corrections, which contribute to orders ${\cal{O}}(\alpha_{em}^3)$ and ${\cal{O}}(\alpha_{em}^2 (m_d - m_u) / \Lambda_{QCD})$, respectively.
The lattice calculation of the IB corrections is performed within the RM123 approach~\cite{deDivitiis:2011eh,deDivitiis:2013xla}, which consists in the expansion of the path integral in powers of the $u$- and $d$-quark mass difference ($m_d - m_u$) and of the em coupling $\alpha_{em}$.
The quenched-QED (qQED) approximation, which treats the dynamical quarks as electrically neutral particles, is adopted and quark-disconnected contractions are not yet included because of the large statistical fluctuations of the corresponding signals.

\section{The HVP contribution to the lepton anomalous magnetic moment}
\label{sec:HVP}

The HVP contribution $a_\ell^{\rm HVP}$ to the lepton anomalous magnetic moment can be calculated by adopting the time-momentum representation~\cite{Bernecker:2011gh}
\be
    a_\ell^{\rm HVP} = 4 \alpha_{em}^2 \int_0^\infty dt ~ K_\ell(t) V(t) ~ ,
    \label{eq:amu_t}
\ee
where the kernel function $K_\ell(t)$ is given by
\be
    K_\ell(t) = \frac{4}{m_\ell^2} \int_0^\infty d\omega ~ \frac{1}{\sqrt{4 + \omega^2}} ~ 
                      \left( \frac{\sqrt{4 + \omega^2} - \omega}{\sqrt{4 + \omega^2} + \omega} \right)^2 
                      \left[ \frac{\mbox{cos}(\omega m_\ell t) - 1} {\omega^2} + \frac{1}{2} m_\ell^2 t^2 \right] 
    \label{eq:kernel}
\ee
with $m_\ell$ being the lepton mass.
In Eq.~(\ref{eq:amu_t}) the quantity $V(t)$ is the vector current-current Euclidean correlator defined as
\be
    V(t) \equiv -\frac{1}{3} \sum_{i=1,2,3} \int d\vec{x} ~ \langle J_i(\vec{x}, t) J_i(0) \rangle ~ ,
    \label{eq:VV}
\ee
where
 \be
     J_\mu(x) \equiv \sum_{f = u, d, s, c, ...} J_\mu^f(x) =  \sum_{f = u, d, s, c, ...} q_f ~ \overline{\psi}_f(x) \gamma_\mu \psi_f(x)
     \label{eq:Jem}
 \ee
is the em current with $q_f$ being the electric charge of the quark with flavor $f$ in units of the electron charge $e$, while $\langle ... \rangle$ means the average of the $T$-product over gluon and quark fields. 
We consider only the quark-connected HVP contributions, so that each quark flavor $f$ contributes separately.

According to the RM123 method, for each quark flavor $f$ the vector correlator $V_f(t)$ is expanded into the sum of a lowest-order contribution $V_f^{LO}(t)$, evaluated in isospin-symmetric QCD (i.e.~$m_u = m_d$ and $\alpha_{em} = 0$), and of a correction $V_f^{NLO}(t)$, computed to leading order in the small parameters $(m_d - m_u) / \Lambda_{QCD}$ and $\alpha_{em}$.
The separation between the isosymmetric QCD and the IB contributions is performed following the Gasser-Rusetsky-Scimemi (GRS) prescription~\cite{Gasser:2003hk}.
Thus, $a_\ell^{\rm HVP}$ is given by the sum of six terms
\bea
    a_\ell^{\rm HVP} & = & a_\ell^{\rm HVP, LO}(ud) +  a_\ell^{\rm HVP, LO}(s) + a_\ell^{\rm HVP, LO}(c) \nonumber \\
                               & + & a_\ell^{\rm HVP, NLO}(ud) + a_\ell^{\rm HVP, NLO}(s) + a_\ell^{\rm HVP, NLO}(c) ~ , ~
    \label{eq:amuf}
\eea
where the contribution of the heavier $b$-quark is known to be negligible with respect to the current level of the uncertainties.

In the case of the muon the strange and charm contributions $a_\mu^{\rm HVP, LO}(s)$ and $a_\mu^{\rm HVP, LO}(c)$ have been determined in Ref.~\cite{Giusti:2017jof}, the light-quark contribution $a_\mu^{\rm HVP, LO}(ud)$ in Ref.~\cite{Giusti:2018mdh} and the three IB corrections $a_\mu^{\rm HVP, NLO}(ud)$, $a_\mu^{\rm HVP, NLO}(s)$ and $a_\mu^{\rm HVP, NLO}(c)$ in Ref.~\cite{Giusti:2019xct}, where a non-perturbative determination of the QED corrections to the relevant renormalization constants is also included.

For the electron and the tau each of the six contributions are evaluated in the same way as the corresponding term for the muon, but using the appropriate kernel $K_\ell(t)$.
The extrapolation to the physical pion point and the continuum and infinite volume limits are performed using the same fitting formulas adopted in Refs.~\cite{Giusti:2018mdh,Giusti:2017jof,Giusti:2019xct}.

We do not provide here the details of the analysis for the electron and the tau.
We limit ourselves to mention that the corrections due to finite volume effects (FVEs) are important in the case of the light-quark LO contribution $a_\ell^{\rm HVP, LO}(ud)$, which in turn dominates $a_\ell^{\rm HVP}$.
In Ref.~\cite{Giusti:2018mdh} an analytic representation of the light-quark vector correlator $V_{ud}^{LO}(t)$, based on quark-hadron duality~\cite{SVZ} at small time distances and on the two-pion finite-volume energy spectrum~\cite{Luscher:1990ux,Lellouch:2000pv} at intermediate and large time distances, was successfully applied to the lattice data, allowing a direct lattice estimate of FVEs. It turned out that FVEs are quite important and much larger than the prediction of Chiral Perturbation Theory (ChPT) at NLO (see Fig.~19 of Ref.~\cite{Giusti:2018mdh} and also Table~1 of Ref.~\cite{Giusti:2019dmu} for the case of the muon).
Indeed, FVEs corresponding to NLO ChPT are known to ignore the interaction among the two pions~\cite{Aubin:2015rzx}.

An important finding of the analysis of Ref.~\cite{Giusti:2018mdh} is that the ``dual + $\pi \pi$'' representation of $V_{ud}^{LO}(t)$ can be extrapolated to the physical pion mass and to the continuum and infinite volume limits.
Therefore, an independent determination of $a_\mu^{\rm HVP, LO}(ud)$ at the physical point was obtained and shown to agree nicely with the direct chiral extrapolation of the lattice data. 
The same holds as well in the case of $a_e^{\rm HVP, LO}(ud)$ and $a_\tau^{\rm HVP, LO}(ud)$.
Thus, within the current level of precision our determination of $a_\ell^{\rm HVP, LO}(ud)$ do not suffer from the lack of simulations at the physical pion mass. 
More important than that is to avoid the use of NLO ChPT for correcting the FVEs.

Before closing this section we mention that the MUonE experiment~\cite{CarloniCalame:2017jfa} aims at determining $a_\mu^{\rm HVP}$ by measuring the running of $\alpha_{em}(q^2)$ for space-like values of the squared four-momentum transfer $q^2$ using a muon beam on a fixed electron target. 
There is, however, a kinematical region not covered by the MUonE experiment and the corresponding contribution to $a_\mu^{\rm HVP}$ needs to be estimated using either $e^+ e^-$ data or lattice QCD simulations.
The contribution to $a_\mu^{\rm HVP}$ not covered by the MUonE experiment, which will be referred to as $a_{MUonE}^{\rm HVP}$, is given by Eq.~(\ref{eq:amu_t}) with the kernel $K_{\ell = \mu}(t)$ replaced by
\be
    K_{MUonE}(t) = \frac{4}{m_\mu^2} \int_{\overline{\omega}}^\infty d\omega ~ \frac{1}{\sqrt{4 + \omega^2}} ~ 
                              \left( \frac{\sqrt{4 + \omega^2} - \omega}{\sqrt{4 + \omega^2} + \omega} \right)^2 
                              \left[ \frac{\mbox{cos}(\omega m_\mu t) - 1} {\omega^2} + \frac{1}{2} m_\mu^2 t^2 \right] ~ , ~
    \label{eq:kernel_MUONE}
\ee
where $\overline{\omega} = 0.93 / \sqrt{1 - 0.93} \simeq 3.5$.

In what follows the four quantitites $a_\ell^{\rm HVP, LO}(ud)$ with $\ell = \{e, \mu, \tau, MUonE\}$ will be evaluated using the ``dual + $\pi \pi$'' representation $[V_{ud}^{LO}(t)]^{phys}$ extrapolated directly to the physical point.

\begin{figure}[htb!]
\begin{center}
\includegraphics[scale=0.75]{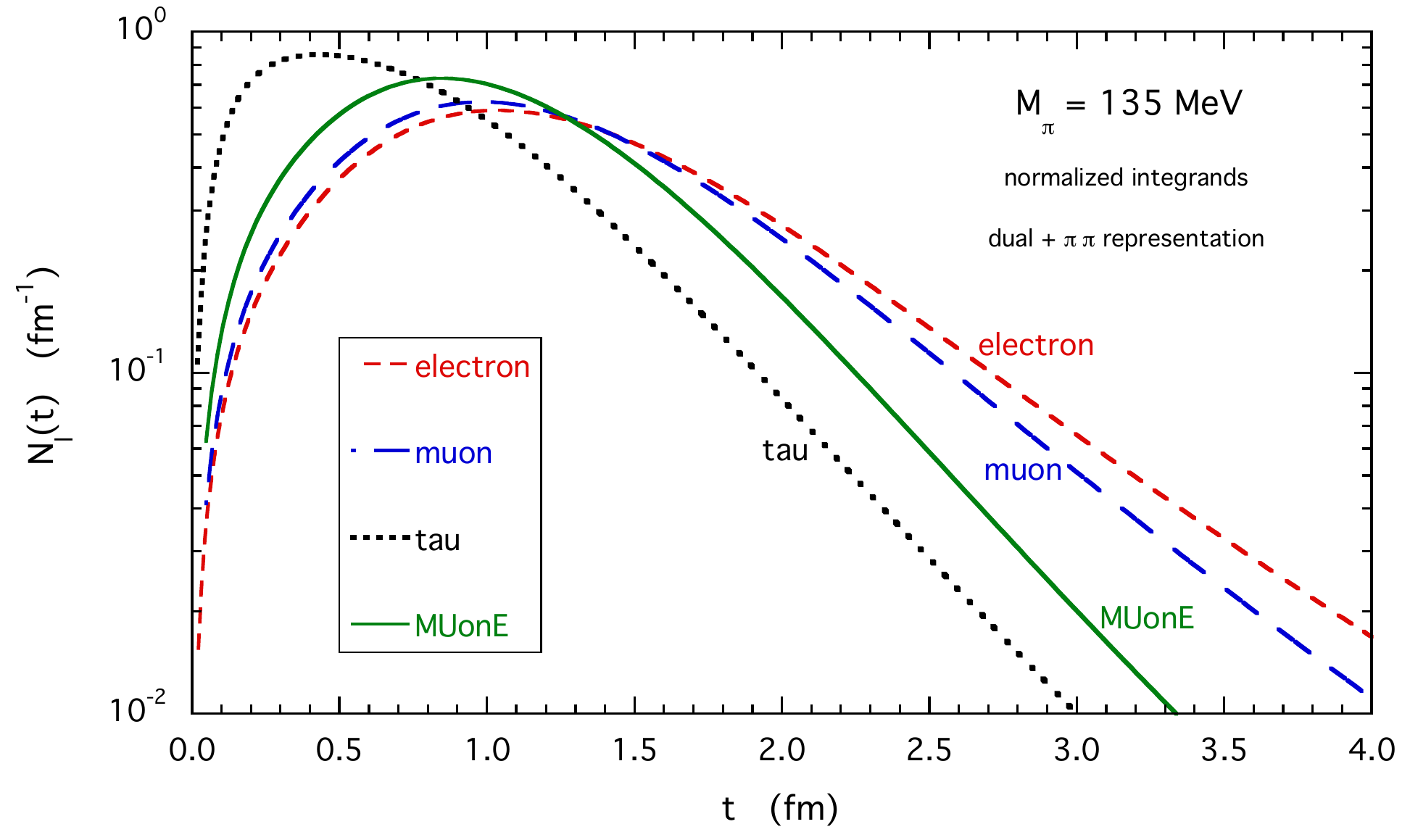}
\end{center}
\vspace{-0.5cm}
\caption{\it \small Time behavior of the normalized integrands $N_\ell(t)$, given by Eq.~(\ref{eq:kernels_norm}), for $\ell = \{e, \mu, \tau, MUonE\}$.}
\label{fig:kernels}
\end{figure}

In the case of the LO light-quark contribution the time behavior of the normalized integrands of Eq.~(\ref{eq:amu_t}) , namely
\be
    N_\ell(t) \equiv K_\ell(t) \cdot \left[ V_{ud}^{LO}(t) \right]^{phys} \Big/ \int_0^\infty dt K_\ell(t) \cdot \left[ V_{ud}^{LO}(t) \right]^{phys} ~ , ~ 
    \qquad \ell = \{e, \mu, \tau, MUonE\}
    \label{eq:kernels_norm}
\ee
is shown in Fig.~\ref{fig:kernels}, which illustrates the relative weights of the various regions of time distances $t$ contributing to $a_\ell^{\rm HVP, LO}(ud)$ for the different leptons, since by definition $\int_0^\infty dt N_\ell(t) = 1$.

\section{Results for $a_{e, \mu, \tau}^{HVP}$ and $a_{MUonE}^{HVP}$}
\label{sec:results}

The results obtained at the physical point for each of the six terms appearing in Eq.~(\ref{eq:amuf}) are shown in Tables~\ref{tab:LO} and \ref{tab:NLO} together with their sum over the four flavors considered.
The uncertainties represent the sum in quadrature of various sources of errors, namely statistical, fitting procedure, input parameters, discretization, FVEs and chiral extrapolation (see for details Ref.~\cite{Giusti:2018mdh}).
In Table~\ref{tab:NLO} an estimate of the error due to the qQED approximation is also included~\cite{Giusti:2019xct}.
\begin{table}[htb!]
    \renewcommand{\arraystretch}{1.25}
    \small{
    \centering    
     \begin{tabular}{|L|c|c|c|c|}
     \hline
     f & $a_e^{\rm HVP, LO}(f) \cdot 10^{14}$ & $a_\mu^{\rm HVP, LO}(f) \cdot 10^{10}$ & $a_\tau^{\rm HVP, LO}(f) \cdot 10^8$ & $a_{MUonE}^{\rm HVP, LO}(f) \cdot 10^{10}$\\
     \hline \hline
    ud                                   & 170.7 ~ (3.9)   & 629.1 ~ (13.7)        & 273.3 ~ (6.6)  & 81.2 ~ (1.7)\\ \hline
    s                                     & ~~13.5 ~ (0.8) & 53.1 ~ (2.5)            & ~36.2 ~ (1.1)  & ~8.3 ~ (0.4)\\ \hline
    c                                     & ~~~3.5 ~ (0.2) & ~~~14.75 ~ (0.56) & ~25.8 ~ (0.8)  & ~2.8 ~ (0.1)\\ \hline \hline
    udsc                               & 187.7 ~ (4.0)   & 697.0 ~ (13.9)        & 335.3 ~ (6.7)  & 92.3 ~ (1.7)\\ \hline
     \end{tabular}
     \caption{\it \small Results for the (connected) LO terms $a_\ell^{\rm HVP, LO}(f)$ corresponding to the various flavors $f = \{ ud, s, c\}$ and leptons $\ell = \{e, \mu, \tau, MUonE\}$. For the light-quark contributions the ``dual + $\pi \pi$'' representation $[V_{ud}^{LO}(t)]^{phys}$ extrapolated to the physical point is adopted. The values of $a_{MUonE}^{\rm HVP, LO}(s)$ and $a_{MUonE}^{\rm HVP, LO}(c)$ have been determined in Ref.~\cite{Giusti:2019hoy}. The last row contains the sum over the four flavors considered.}
     \label{tab:LO}
     }
    \renewcommand{\arraystretch}{1.0}
\end{table}

\begin{table}[htb!]
   \renewcommand{\arraystretch}{1.25}
   \small{
   \centering
   \begin{tabular}{|L|c|c|c|c|}
   \hline
   f & $a_e^{\rm HVP, NLO}(f) \cdot 10^{14}$ & $a_\mu^{\rm HVP, NLO}(f) \cdot 10^{10}$ & $a_\tau^{\rm HVP, NLO}(f) \cdot 10^8$ & $a_{MUonE}^{\rm HVP, NLO}(f) \cdot 10^{10}$\\
   \hline \hline
   ud             & 1.9 ~ (0.8)            & 7.1 ~ (2.5)               & 3.0 ~ (1.1)         & 0.9 ~ (0.3)\\ \hline
   s               & -0.002 ~ (0.001)   & -0.0053 ~ (0.0033)  & 0.001 ~ (0.002) & -0.0005 ~ (0.0004)\\ \hline
   c               & ~0.004 ~ (0.001)  & ~0.0182 ~ (0.0036) & 0.032 ~ (0.006) & ~0.0034 ~ (0.0007)\\ \hline \hline
   udsc         & 1.9 ~ (1.0)            & 7.1 ~ (2.9)                & 3.0 ~ (1.3)        & 0.9 ~ (0.3) \\ \hline
   \end{tabular}
    \caption{\it \small The same as in Table~\ref{tab:LO}, but for the NLO terms $a_\ell^{\rm HVP, NLO}(f)$ corresponding to strong and em IB effects. The values in the last column have been determined in Ref.~\cite{Giusti:2019hoy}. In the last row the uncertainties include also an estimate of the error due to the qQED approximation~\cite{Giusti:2019xct}.}
    \label{tab:NLO}
    }
    \renewcommand{\arraystretch}{1.0}
\end{table}
In Ref.~\cite{Giusti:2018mdh}, in the case of the muon, we estimated the contribution of the quark disconnected diagrams to be equal to $a_\mu^{\rm HVP}(disconn.) = -12 ~ (4) \cdot 10^{-10}$, obtained using the findings of Refs.~\cite{Borsanyi:2017zdw,Blum:2018mom}.
In the case of the electron and the tau we adopt directly the findings of Ref.~\cite{Borsanyi:2017zdw}, namely  $a_e^{\rm HVP}(disconn.) = -3.8 ~ (0.4) \cdot 10^{-14}$ and $a_\tau^{\rm HVP}(disconn.) = -2.4 ~ (0.3) \cdot 10^{-8}$.
For MuonE we adopt the following strategy.
First, we consider the ratio of disconnected over connected contributions in the case of the muon, namely $-12 \, (4) / \, 697.0 \, (13.9) = -0.0172 ~ (57)$.
Then, the same value of the ratio is assumed to hold as well for MUonE, implying that $a_{MUonE}^{\rm HVP}(disconn.) = -1.6 ~ (0.5) \cdot 10^{-10}$.

For conservative purposes we {\it double} the uncertainty of the above estimates of the disconnected diagrams.
Adding all the various contributions we get
\bea
      \label{eq:ae_ETMC}
      a_e^{\rm HVP} & = & 185.8 ~ (4.2) \cdot 10^{-14} ~ , ~ \\[2mm]
       \label{eq:amu_ETMC}
      a_\mu^{\rm HVP} & = & 692.1 ~ (16.3) \cdot 10^{-10} ~ , ~ \\[2mm]
      \label{eq:atau_ETMC}
      a_\tau^{\rm HVP} & = & 335.9 ~ (6.9) \cdot 10^{-8} ~ , ~ \\[2mm]
      \label{eq:aMUONE_ETMC}
      a_{MUonE}^{\rm HVP} & = & ~~\, 91.6 ~ (2.0) \cdot 10^{-10} ~ , ~
\eea
where the muon result (\ref{eq:amu_ETMC}) represents an update of the corresponding ETMC result of Ref.~\cite{Giusti:2018mdh}.

Our results (\ref{eq:ae_ETMC})-(\ref{eq:atau_ETMC}) can be compared with the corresponding ones from the BMW Collaboration~\cite{Borsanyi:2017zdw}: $a_e^{\rm HVP} = 189.3 ~ (6.2) \cdot 10^{-14}$, $a_\mu^{\rm HVP} = 711.1 ~ (18.9) \cdot 10^{-10}$ and $a_\tau^{\rm HVP} = 341.0 ~ (3.3) \cdot 10^{-8}$.
They remarkably agree well with the more precise determinations based on the dispersive analyses of the experimental cross section data for $e^+ e^-$ annihilation into hadrons: $a_e^{\rm HVP} = 184.90 ~ (1.08) \cdot 10^{-14}$~\cite{Jegerlehner:2017zsb}, $a_\mu^{\rm HVP} = 693.9 ~ (4.0) \cdot 10^{-10}$~\cite{Davier:2019can} and $a_\tau^{\rm HVP} = 337.5 ~ (3.7) \cdot 10^{-8}$~\cite{Eidelman:2007sb}

Notice that the uncertainty of our result (\ref{eq:aMUONE_ETMC}) is almost coinciding with the statistical uncertainty ($\sim 2 \cdot 10^{-10}$) expected in the MUonE experiment for the measured contribution $[a_\mu^{\rm HVP} - a_{MUonE}^{\rm HVP}]$ after two years of data taking at the CERN North Area~\cite{CarloniCalame:2017jfa}. 

Before closing this Section we take advantage of the well known fact that $a_e^{\rm HVP}$ is proportional to the slope $\Pi_1$ of the leading HVP function at vanishing photon virtuality, namely $a_e^{\rm HVP} = (4/3) \alpha_{em}^2 m_e^2 \Pi_1$ with $\Pi_1 \equiv (1/12) \int_0^\infty dt ~ t^4 V(t)$.
Our result (\ref{eq:ae_ETMC}) corresponds to
\be
    \Pi_1 = 0.1002 ~ (23) ~ \mbox{GeV}^{-2} ~ ,
\ee
which agrees nicely with the recent results $\Pi_1 = 0.1000 ~ (30)$ GeV$^{-2}$ and $\Pi_1 = 0.1000 ~ (23)$ GeV$^{-2}$, obtained by the BMW and FNAL/HPQCD/MILC Collaborations in Refs.~\cite{Borsanyi:2016lpl,Davies:2019efs}, respectively.

\end{document}